\documentclass[aps,prl,reprint,amsmath,amssymb,superscriptaddress,floatfix,nobibnotes]{revtex4-2}

\usepackage[T1]{fontenc}
\usepackage[utf8]{inputenc}
\usepackage{lmodern}
\usepackage{graphicx}
\usepackage{bm}
\usepackage{microtype}
\usepackage{indentfirst}
\usepackage{xcolor}
\usepackage[hidelinks]{hyperref}
\hypersetup{pdfborder={0 0 0}}
\AtBeginDocument{%
  \setlength{\abovedisplayskip}{5pt plus 1pt minus 1pt}%
  \setlength{\belowdisplayskip}{5pt plus 1pt minus 1pt}%
  \setlength{\abovedisplayshortskip}{3pt plus 1pt minus 1pt}%
  \setlength{\belowdisplayshortskip}{4pt plus 1pt minus 1pt}%
  \setlength{\jot}{2.5pt}%
  \setlength{\textfloatsep}{8pt plus 1pt minus 1pt}%
  \setlength{\floatsep}{7pt plus 1pt minus 1pt}%
  \setlength{\intextsep}{7pt plus 1pt minus 1pt}%
  \setlength{\parskip}{0pt}%
  \setlength{\parindent}{1em}%
  \setlength{\abovecaptionskip}{4pt}%
  \setlength{\belowcaptionskip}{0pt}%
}

\newcommand{\CGT}{\mathrm{Cr_2Ge_2Te_6}}
\newcommand{\smartinclude}[2][]{%
  \IfFileExists{#2.pdf}{\includegraphics[#1]{#2.pdf}}{%
  \IfFileExists{#2.png}{\includegraphics[#1]{#2.png}}{%
  \IfFileExists{#2.jpg}{\includegraphics[#1]{#2.jpg}}{%
  \IfFileExists{#2.jpeg}{\includegraphics[#1]{#2.jpeg}}{%
    \fbox{\parbox[c][0.12\textheight][c]{0.92\linewidth}{\centering\small Figure file placeholder\\[2pt]\texttt{#2.png / #2.jpg}}}%
  }}}}%
}

\begin{document}

\preprint{APS/XXX-Q}
\title{Nonlinear Magneto-Optical Probing of Time-Reversal Symmetry Breaking}

\author{Chen Qian}
\affiliation{Institute of Ultrafast Optical Physics, Department of Applied Physics, Nanjing University of Science and Technology, Nanjing 210094, P. R. China}
\affiliation{Institute of Physics, Chinese Academy of Sciences, Beijing National Laboratory for Condensed Matter Physics, Beijing 100190, P. R. China}
\author{Ziwen Wang}
\affiliation{Institute of Ultrafast Optical Physics, Department of Applied Physics, Nanjing University of Science and Technology, Nanjing 210094, P. R. China}
\author{Shicheng Jiang}
\affiliation{Institute of Ultrafast Optical Physics, Department of Applied Physics, Nanjing University of Science and Technology, Nanjing 210094, P. R. China}
\author{Hui Li}
\affiliation{Institute of Ultrafast Optical Physics, Department of Applied Physics, Nanjing University of Science and Technology, Nanjing 210094, P. R. China}
\author{Kun Zhao}
\affiliation{Institute of Ultrafast Optical Physics, Department of Applied Physics, Nanjing University of Science and Technology, Nanjing 210094, P. R. China}
\author{Chao Yu}
\email{chaoyu@njust.edu.cn}\affiliation{Institute of Ultrafast Optical Physics, Department of Applied Physics, Nanjing University of Science and Technology, Nanjing 210094, P. R. China}
\author{Yuan Wan}
\affiliation{Institute of Physics, Chinese Academy of Sciences, Beijing National Laboratory for Condensed Matter Physics, Beijing 100190, P. R. China}
\author{Hongming Weng}
\email{hmweng@iphy.ac.cn}\affiliation{Institute of Physics, Chinese Academy of Sciences, Beijing National Laboratory for Condensed Matter Physics, Beijing 100190, P. R. China}
\affiliation{School of Physics, University of Chinese Academy of Sciences, Beijing 100049, P. R. China}
\affiliation{Songshan Lake Materials Laboratory, Dongguan, Guangdong 523808, P. R. China}
\author{Ruifeng Lu}
\email{rflu@njust.edu.cn}
\affiliation{Institute of Ultrafast Optical Physics,
Department of Applied Physics,
Nanjing University of Science and Technology,
Nanjing 210094, P. R. China}

\date{\today}

\begin{abstract}
Solid-state harmonic generation provides a nonlinear probe of symmetries
encoded in electronic wave functions. In the subgap and weak-injection
regime, time reversal pairs the harmonic responses driven by fields of
opposite ellipticity, strongly suppressing elliptical dichroism in
time-reversal-symmetric crystals. We show that, in a magnetic crystal,
spin-orbit coupling transfers time-reversal-symmetry breaking from the
spin sector to the orbital wave functions and lifts this pairing through
the geometric phases of the electric-dipole current.
Semiconductor-Bloch-equation calculations for centrosymmetric bilayer
$\mathrm{Cr_2Ge_2Te_6}$ predict pronounced third-harmonic elliptical
dichroism that reverses with the magnetization. Under linearly polarized
driving, SOC-induced geometric-phase accumulation generates a nonlinear
transverse current and strongly enhances the harmonic rotation and
ellipticity. These results identify the geometric phase as a key microscopic
contribution to the nonlinear magneto-optical response. This work
establishes helicity-resolved harmonic emission and nonlinear polarimetry
as complementary probes of spin-orbit-coupled magnetic order.
\end{abstract}

\maketitle

Harmonic generation in solids provides a nonlinear,
momentum-space probe of crystal symmetry, band topology, wave-function
geometry, and electronic correlations
\cite{Alon1998PRL,Neufeld2019NatCommun,Qian2024PRB,
Morimoto2016SciAdv,Lv2021NatCommun,Silva2019NatPhoton,
Qian2022PRX,Murakami2018PRL,Silva2018NatPhoton,Shao2022PRL}.
The polarization and selection rules of solid-state harmonics encode the
spatiotemporal symmetries of both the crystal and the driving field, while
helicity-resolved excitation can selectively access spin, valley, and
chiral electronic degrees of freedom
\cite{Xiao2010RMP,Xiao2012PRL,Neufeld2019PRX,Yu2016PRB,
Ma2017NatPhys}. High-order and ultrafast optical measurements have also
emerged as probes of spin-polarized dynamics and magnetic order
\cite{Zhang2018NatCommun,Willems2015PRB,Fan2015PNAS,
LaOVorakiat2009PRL}. A central question nevertheless remains unresolved:
What constraint does time-reversal symmetry (TRS) impose on strong-field harmonic emission, and how is this constraint lifted by
magnetic order? This question is nontrivial because magnetic order resides
primarily in the spin sector, whereas the leading strong-field optical
response is generally governed by electric-dipole currents in the orbital
sector. Spin polarization alone therefore need not produce
helicity-dependent electric-dipole emission. A microscopic mechanism is
required to imprint the broken TRS of the spin sector
onto the orbital wave functions and their light-driven dynamics.

Established magneto-optical probes, including magnetic circular dichroism
and the Faraday and Kerr effects, detect the coupling of magnetic order to
optical transitions
\cite{Gong2017Nature,Nemec2018NatPhys,Huang2017Nature,
Sato2022FrontPhys}. Magnetic states may additionally be prepared, switched,
or calibrated by external fields or ultrafast excitation
\cite{Liu2016JACS,Willems2015PRB,Fan2015PNAS,
LaOVorakiat2009PRL}, although such perturbations may be undesirable for
delicate spin textures and nearly compensated magnetic phases.
Magnetic second-harmonic generation (SHG) provides a powerful nonlinear
and symmetry-sensitive alternative
\cite{Sun2019Nature,Fiebig2005JOSAB,Sipe2000PRB,
Fiebig1994PRL,Pi2023QuantumFrontiers,Wu2024NPJCompMat}.
Its bulk electric-dipole contribution, however, is forbidden in
centrosymmetric crystals and may be obscured by nonmagnetic
crystallographic contributions when inversion symmetry is absent
\cite{Fiebig2005JOSAB}. Odd-order harmonic generation remains
electric-dipole allowed in centrosymmetric systems and can therefore
provide a complementary channel for detecting magnetic symmetry. Recent
work has demonstrated a strong-field enhancement of third-harmonic (H3)
magneto-optical rotation under optically induced time-reversal-symmetry
breaking \cite{Friedrich2026NatPhoton}. This advance raises a distinct but
complementary question: Can equilibrium magnetic order be identified
through a symmetry constraint on helicity-resolved harmonic emission,
rather than solely through the magnitude of a conventional weak-probe
rotation?

We first establish a symmetry-based helicity-pairing criterion for strong-field harmonic generation. In the subgap and weak-injection regime, where nearly reversible interband polarization dominates the optical response, time reversal pairs the harmonic signals driven by fields of opposite ellipticity. Consequently, time-reversal-symmetric crystals exhibit strongly suppressed harmonic elliptical dichroism (ED), in agreement with previous results for time-reversal-symmetric semiconductors.
\cite{TancogneDejean2017NatCommun,Klemke2019NatCommun}. In a magnetic
crystal, spin-orbit coupling (SOC) transfers time-reversal-symmetry breaking from the spin sector to
the orbital wave functions, removes this pairing constraint, and generates a
magnetization-dependent contribution through the geometric phases of the
electric-dipole current. Semiconductor-Bloch-equation calculations for
centrosymmetric bilayer $\mathrm{Cr_2Ge_2Te_6}$ predict pronounced ED in the
H3, while reversing both the magnetization and the driving
ellipticity restores the time-reversal-related response. Under linearly polarized driving, the same mechanism
generates a nonlinear transverse current and strongly enhances the
harmonic polarization rotation and ellipticity. These results
identify helicity-resolved harmonic emission as a nonlinear magneto-optical
signature of time-reversal-symmetry breaking that can be measured without an
external magnetic field applied during optical readout.

\begin{figure}[t]
\centering
\smartinclude[width=0.95\columnwidth]{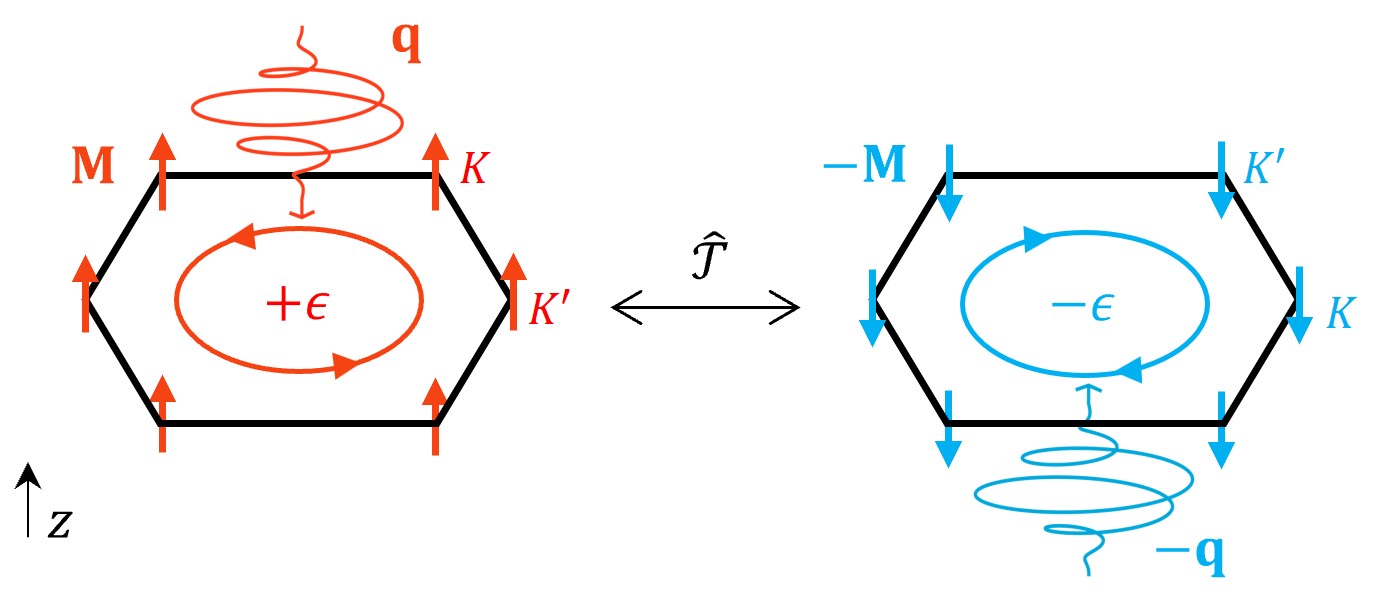}
\caption{Under time reversal $\hat{\mathcal{T}}$, the crystal magnetization
$\mathbf{M}$ and the wave vector $\mathbf{q}$ of the laser field are
reversed, while its ellipticity changes sign.
Here, $+\epsilon$ and $-\epsilon$ denote left- and right-handed
polarizations, respectively, defined with respect to the fixed
laboratory $z$ axis. The corresponding time-reversed high-symmetry
$\mathbf{k}$ points are marked.}
\label{fig:1}
\end{figure}

The symmetry principle is illustrated in Fig.~\ref{fig:1}. Time reversal maps an
in-plane elliptically polarized drive onto one with the opposite ellipticity $\epsilon$, \(\hat{\mathcal{T}}:
\mathbf{E}(\epsilon,t)\longrightarrow\mathbf{E}(-\epsilon,-t)=\mathbf{E}(\epsilon,t)\)
and transforms the corresponding crystal-momentum trajectory as
\(\mathbf{k}(\epsilon,t)\longrightarrow-\mathbf{k}(-\epsilon,-t)=\mathbf{k}(\epsilon,t).\)
Apply $\hat{\mathcal{T}}$ to a laser-crystal system, the currents generated by these paired drives satisfy
\(\mathbf{J}(\epsilon,\mathbf{M};t)\longrightarrow-\mathbf{J}(-\epsilon,-\mathbf{M};-t)=\mathbf{J}(\epsilon,\mathbf{M};t),\)
up to the appropriate mapping between time-reversed bands. 
The corresponding Fourier amplitudes thus have identical moduli, implying
\(
I(\epsilon,\mathbf{M};\omega)=I(-\epsilon,-\mathbf{M};\omega).
\)
We quantify the response by the elliptic dichroism
\(
\mathrm{ED}=
\frac{I(\epsilon,\omega)-I(-\epsilon,\omega)}
     {I(\epsilon,\omega)+I(-\epsilon,\omega)},
\)
defined for opposite ellipticities of the driving field. Under the exactly
reversible dynamics considered above, \(\mathrm{ED}\) vanishes. Carrier injection,
dissipation, heating, or temporal asymmetry of the pulse may break this condition
and generate a nonmagnetic dichroic background. We therefore work in the subgap and
weak-injection regime, where the carrier transfer is below \(0.5\%\) and the
residual nonmagnetic ED is negligible.

Within the electric-dipole approximation, the
polarization current between any two independent bands $n$, $m$ can be written as
(atomic units are used throughout unless otherwise stated)
\vspace{-2pt}
\begin{equation}
\begin{aligned}
\mathbf{J}_{nm}[\mathbf{k}(t),t]
&=-\rho_{nm}[\mathbf{k}(t),t] \, \mathbf{p}_{nm}[\mathbf{k}(t)] \\
&=-\int_{-\infty}^{t} dt'\,\varepsilon_{nm}[\mathbf{k}(t)]
\left|\mathbf{d}_{nm}[\mathbf{k}(t)]\right| \\
&\quad\times\left\{\mathbf{E}(t')\!\cdot\!\left|\mathbf{d}_{mn}[\mathbf{k}(t')]\right|\right\}
 f_{nm}[\mathbf{k}(t'),t'] \\
&\quad\times\exp [iS_{nm}(\mathbf{k},t,t')].
\end{aligned}
\label{eq:main-current}
\end{equation}
Here, $\rho_{nm}$, $\mathbf{p}_{nm}$, $\varepsilon_{nm}$, and $\mathbf{d}_{nm}$ denote the
density-matrix element, momentum matrix element, band-energy difference,
and transition dipole moment, respectively, while
$f_{nm}=\rho_{nn}-\rho_{mm}$ is the population difference. \(S_{nm}\) is the phase of the
polarization current, which consists of the dynamical phase, the shift
phase and the transition-dipole phase difference (\(\Delta\mathrm{TDP}\)), as
\begin{equation}
\begin{aligned}
S_{nm}(\mathbf{k},t,t')
&=\int_{t'}^{t}\varepsilon_{nm}[\mathbf{k}(\tau)]\,d\tau \\
&\quad+\int_{t'}^{t}\mathbf{E}(\tau)\!\cdot\!\mathbf{R}_{nm}[\mathbf{k}(\tau)]\,d\tau \\
&\quad+S_{\Delta\mathrm{TDP}}[\mathbf{k}(t)].
\end{aligned}
\label{eq:phase}
\end{equation}
\(\mathbf{R}_{nm}\) is the shift vector, \(S_{\Delta\mathrm{TDP}}\) is
the phase difference between transition dipole moments along the photocurrent
and the laser field. The latter two terms are collectively referred to as
the geometric-phase contributions to the polarization current. It has
been shown that the current phase \(S_{nm}\) reverses sign under TRS
\cite{Qian2024PRB}:
\begin{equation}
S_{nm}(\epsilon,\mathbf{M};\mathbf{k},t,t')
=
-S_{\tilde n\tilde m}
(-\epsilon,-\mathbf{M};-\mathbf{k},-t,-t'),
\label{eq:phaserev}
\end{equation}
where \(\tilde{n}\) and \(\tilde{m}\) denote the time-reversed
partners of bands \(n\) and \(m\), respectively.

For subgap driving with negligible population transfer,
$f_{nm}$ remains nearly unchanged and the low-order response is dominated
by the reversible evolution of the interband polarization. The paired polarization current under laser fields of opposite ellipticities satisfies
\begin{equation}
\begin{aligned}
\mathbf{J}_{nm}[\epsilon,\mathbf{M};\mathbf{k}(t),t]
&=-\mathbf{J}_{\tilde{n}\,\tilde{m}}^{*}[-\epsilon,-\mathbf{M};-\mathbf{k}(-t),-t].
\end{aligned}
\label{eq:TR-current}
\end{equation}
The intraband contribution obeys the same time-reversal pairing in the
weak-injection regime.

We illustrate this mechanism in centrosymmetric AB-stacked bilayer
$\mathrm{Cr_2Ge_2Te_6}$, an intrinsic two-dimensional ferromagnet
\cite{Ren2021SciRep}, its long-range ferromagnetic order in the atomically thin limit is enabled by
magnetic anisotropy and SOC. However, magneto-optical Kerr effect
measurements on bilayer
\(\CGT\)~yield
rotation angles of only a few millidegrees \cite{Gong2017Nature}, motivating the development
of complementary probes of its intrinsic magnetism.

\begin{figure}[t]
\centering
\smartinclude[width=\columnwidth]{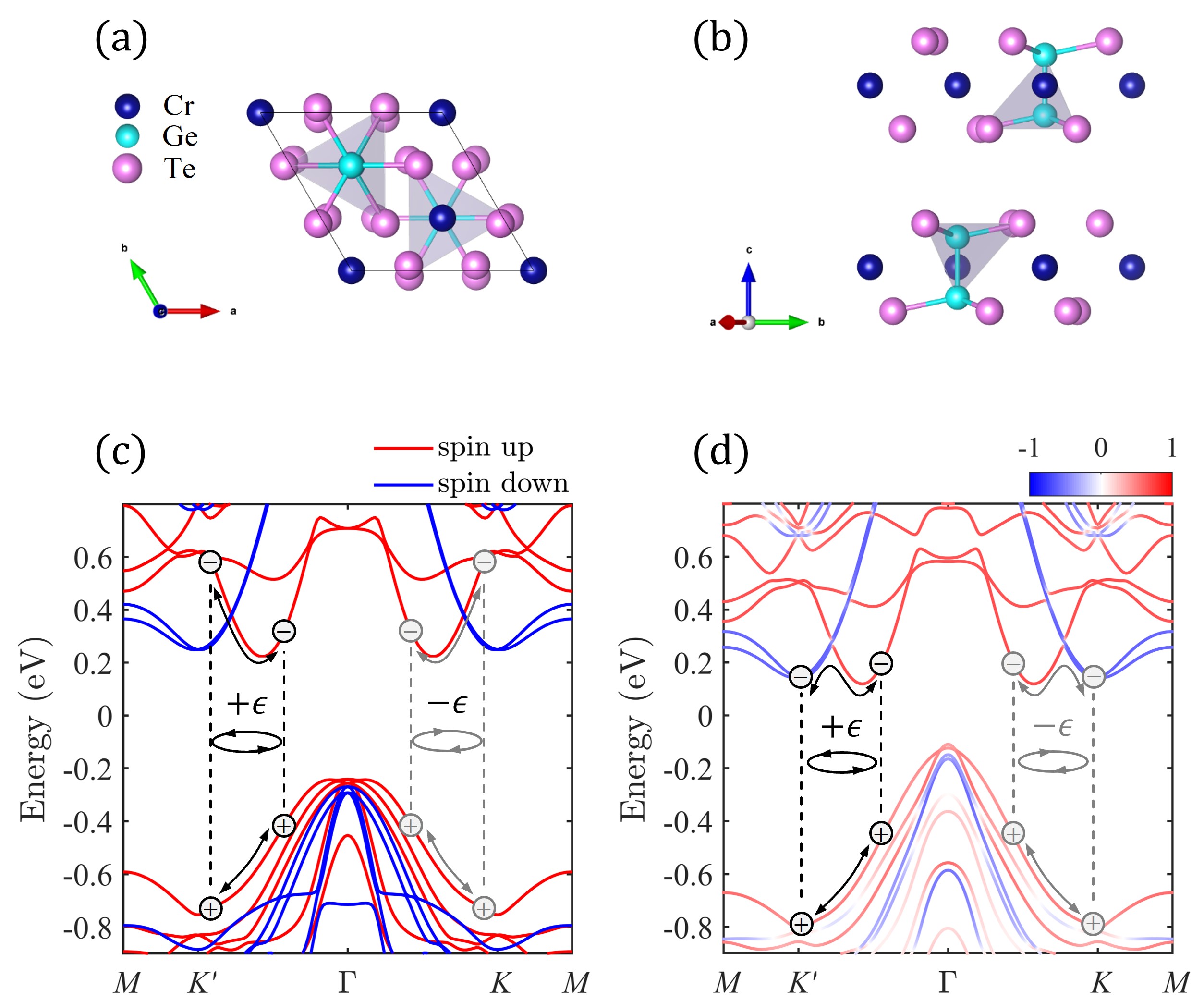}
\caption{Top (a) and side (b) views of the crystal structure of the AB-stacked bilayer $\CGT$, which belongs to the trigonal space group $P\overline{3}$. The rhombic box denotes the primitive cell. Band structures calculated (c) without and (d) with spin-orbit coupling (SOC). The color scale indicates the out-of-plane spin polarization. The black arrows mark the polarization paths of electric dipoles (positive and negative charge pairs) driven by laser fields with opposite ellipticities.}
\label{fig:2}
\end{figure}

The first-principles calculations (see Supplemental Material, Sec. 1 for
computational details) yield the band structure and spin polarization of
\(\CGT\) as shown in
Figs.~\ref{fig:2}(c) and \ref{fig:2}(d). When SOC is not considered, the band structure is
spin-resolved, and there is no coupling between different spin states.
Electric-dipole transitions connect
bands within the same spin sector [Fig.~\ref{fig:2}(c)]. These \(\mathbf{k}\)-resolved dipoles execute a light-driven
quantum evolution, giving rise
to time dependent polarization. SOC mixes the spin
characters and opens avoided crossings [Fig.~\ref{fig:2}(d)], thereby modifying the
orbital wave functions and transition dipole moments sampled along the
field-driven trajectories. These wave-function changes provide the channel
through which magnetic order enters the electric-dipole current. 

We focus on the ED of harmonic spectra induced by the intrinsic
magnetism of materials. A mid-infrared laser field with a wavelength of
\(7.5~\mu\mathrm{m}\), and a full width at half maximum of approximately 5 optical
cycles is used. Over the range of laser intensities employed in this work, the weak-injection condition is satisfied (see Supplemental Fig.~S1). These laser parameters correspond to the multiphoton excitation case. A dephasing time of $T_2=9$ fs is used. In the calculations, we use Wannier90 to construct the
maximally localized Wannier functions to obtain transition dipole
moments \cite{Mostofi2008CPC,Silva2019PRB,Wu2018CPC}, and semiconductor Bloch equations are solved
under the length gauge \cite{Wilhelm2021PRB,Golde2008PRB,Vampa2014PRL,Li2019PRA,Yue2020PRA}. 

To examine the magnetization dependence of the harmonic response, we reverse the crystal magnetization. As shown in Supplemental Fig.~S2, without SOC the harmonic responses for opposite spin polarizations are identical, indicating that the orbital Hamiltonian is unaffected by spin reversal. With SOC, however, the harmonic response becomes spin dependent, demonstrating that SOC transfers time-reversal-symmetry breaking from the spin sector to the orbital wave functions.

\begin{figure}[t]
\centering
\smartinclude[width=0.90\columnwidth]{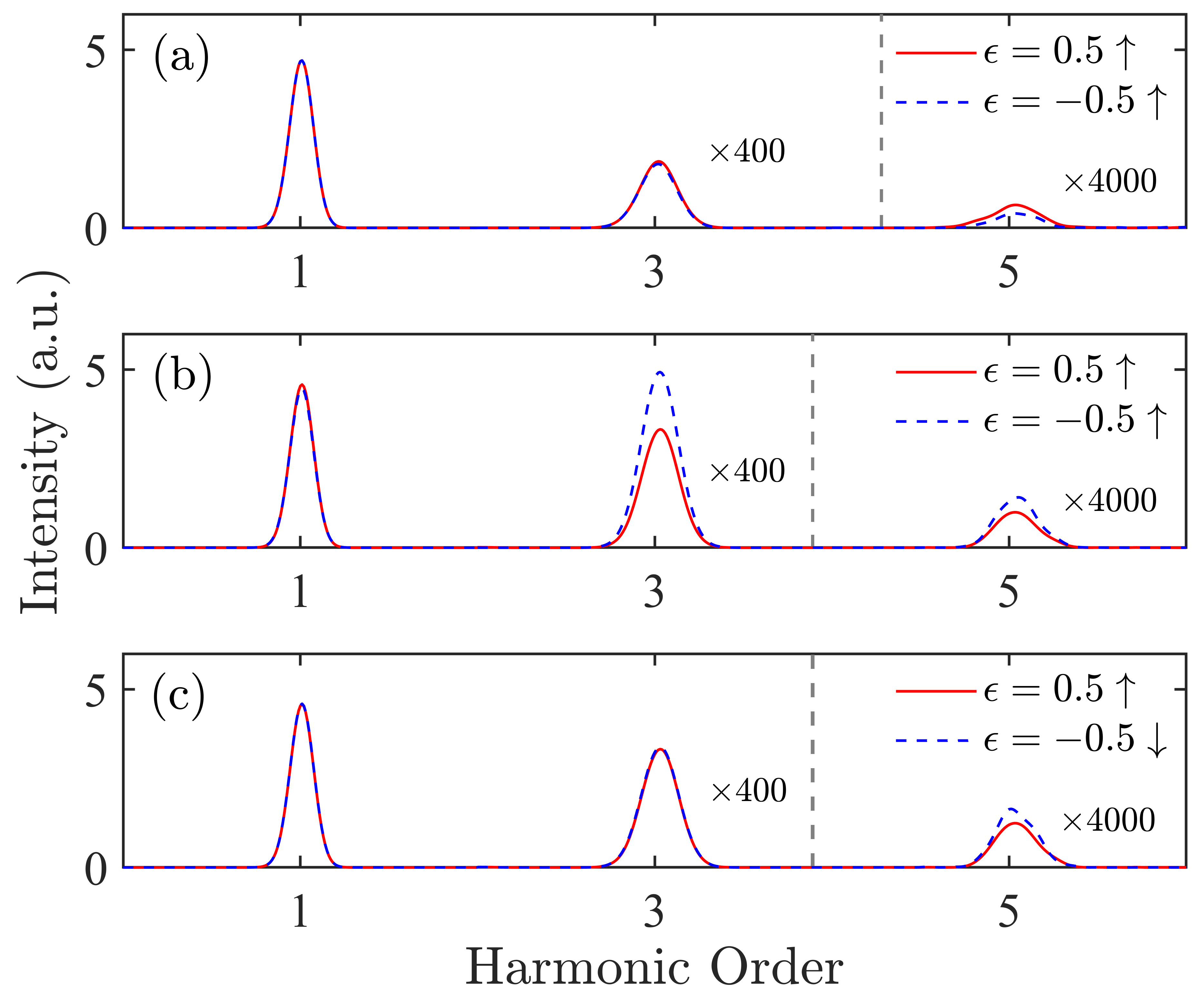}
\caption{Elliptical dichroism of $\CGT$ (a) without SOC and (b, c) with SOC. A laser field with an ellipticity of $\epsilon=\pm0.5$ and a field amplitude of $1.54~\mathrm{MV/cm}$ is used. The gray dotted lines mark the bandgap energy. Arrows denote the spin polarization of the crystal.}
\label{fig:3}
\end{figure}

Figure~\ref{fig:3} shows the harmonic spectra of
\(\CGT\) under laser
fields with opposite ellipticities \(\epsilon = \pm 0.5\). Only odd
harmonics appear due to the inversion symmetry of this material. In the absence of SOC, the subgap ED is negligible. Including SOC produces a pronounced H3 ED of
approximately 0.2 [Fig.~\ref{fig:3}(b)], demonstrating a
strong SOC-enabled nonlinear magneto-optical sensitivity. Upon reversing both the magnetization and the driving ellipticity, the full time-reversal relation is restored, and the ED vanishes again, consistent with Fig.~\ref{fig:3}(c). The intensity dichroism of the fundamental harmonic (H1) is masked by the dominant longitudinal response,
although its polarization state remains sensitive to the magnetic
transverse current, as shown below. The laser field-strength and ellipticity
dependences are provided in Supplemental Figs.~S4 and S5, respectively.

The dichroism originates from the SOC-induced loss of time-reversal pairing
in the polarization-current phase. Inversion symmetry retains the relation
$\varepsilon_{nm}(-\mathbf{k},\mathbf{M})
=\varepsilon_{nm}(\mathbf{k},\mathbf{M})$ and hence the corresponding pairing of
the dynamical phase. 
It does not, however, enforce time-reversal antisymmetry of the shift phase or $\Delta\mathrm{TDP}$ within a fixed magnetic domain, as directly revealed by the calculated shift vectors and $\Delta\mathrm{TDP}$ [Supplemental Fig.~S6]. Microscopically, SOC mixes the magnetic orbital wave functions [Fig.~\ref{fig:2}(d)], thereby breaking the antisymmetric relation between their geometric phases. As the corresponding electric dipoles evolve across momentum space, they accumulate asymmetric geometric phases and generate polarization currents that no longer satisfy the conjugation relation in Eq.~(\ref{eq:TR-current}) [Supplemental Fig.~S3 shows the comparisons of time-domain currents].

\begin{figure}[t]
\centering
\smartinclude[width=\columnwidth]{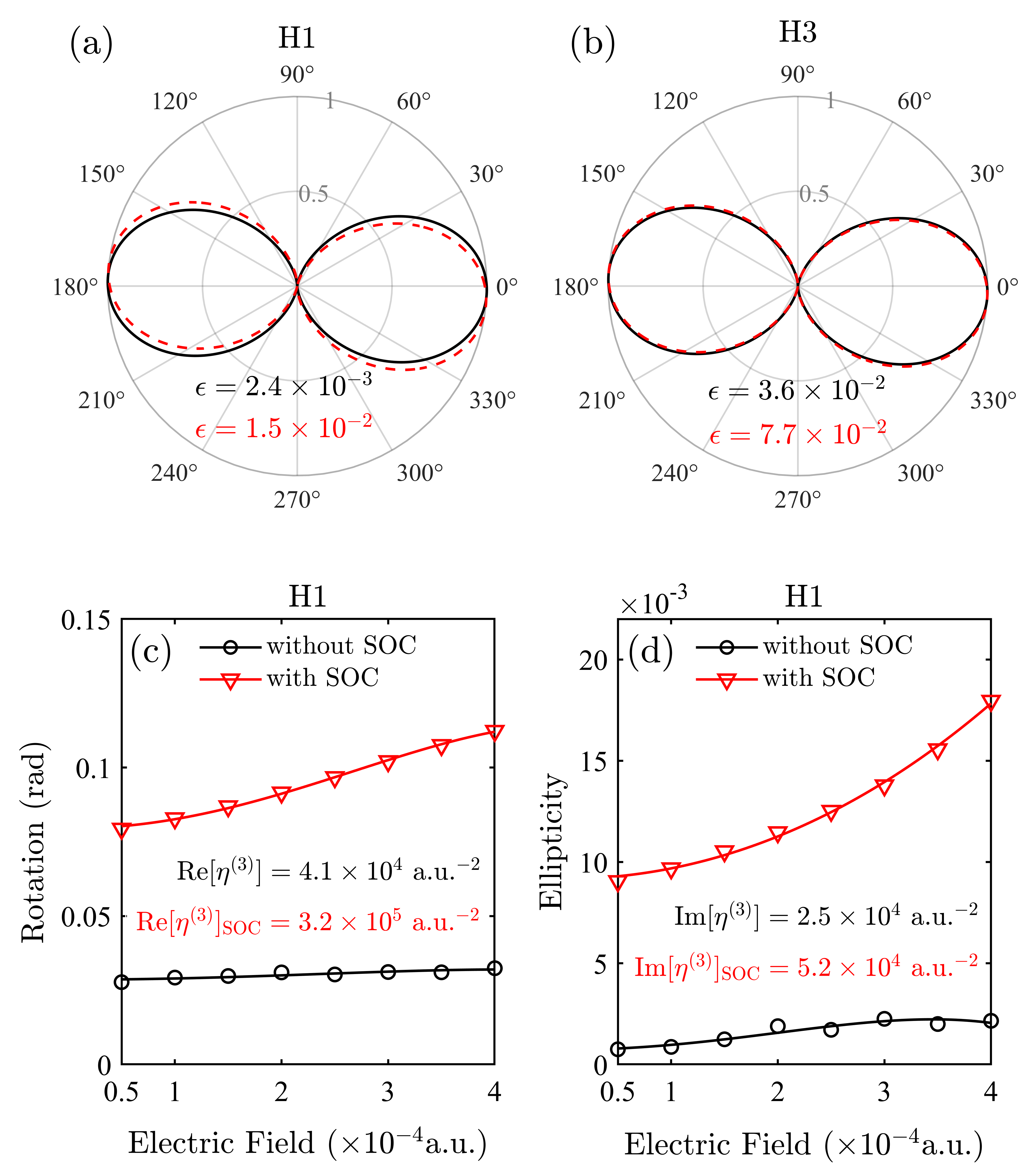}
\caption{Azimuthal distributions of H1 (a) and H3 (b) driven by a linearly polarized laser field with a field amplitude of $1.54~\mathrm{MV/cm}$ ($3\times10^{-4}$ a.u.). Black solid and red dashed lines indicate the cases without and with SOC, respectively. Ellipticities of harmonics are shown with corresponding colors (minor-to-major axis ratio of the harmonic ellipse). Field-dependent polarization rotation (c) and ellipticity (d) of H1. The symbols represent the calculated data, and the solid curves show the fitting results using $\eta=\eta^{(1)}+\eta^{(3)}E^2+\eta^{(5)}E^4$. The extracted quadratic coefficients correspond to $\operatorname{Re}[\eta^{(3)}]$ and $\operatorname{Im}[\eta^{(3)}]$ for the rotation and ellipticity, respectively. The driving laser field is linearly polarized and the orientation is along the $\Gamma K$ direction.}
\label{fig:4}
\end{figure}

In conventional Faraday or Kerr measurements, the magnetic response is
characterized by the rotation and ellipticity acquired by a weak,
linearly polarized probe upon transmission or reflection. Here,
Figs.~\ref{fig:4}(a) and \ref{fig:4}(b) instead resolve the polarization of the H1 and H3 radiation under a strong, linearly polarized drive. Even without SOC, when the electric-dipole current is essentially decoupled from the magnetic order, the harmonics exhibit a small but finite polarization rotation. This residual rotation is nonmagnetic and originates from the intrinsic crystalline anisotropy.

Including SOC enhances the H1 rotation and introduces a nonlinear
dependence on the driving-field strength [Fig.~\ref{fig:4}(c)]. More
strikingly, the H1 ellipticity increases strongly with the field, whereas
only a weak background remains without SOC [Fig.~\ref{fig:4}(d)]. This
background may arise in part from finite dephasing and multiband
interference in the simulations, while the substantially larger
SOC-induced contribution provides the dominant magnetic contrast. At
the field strength used in Fig.~\ref{fig:4}(a), SOC enhances the H1
ellipticity from $2.4\times10^{-3}$ to $1.5\times10^{-2}$.

Microscopically, the finite ellipticity generated by a linearly polarized
field originates from the breakdown of time-reversal pairing between
geometric-phase contributions to the current. In a time-reversal-symmetric
system, the contributions from the paired trajectories
$(\mathbf{k},t)$ and $(-\mathbf{k},-t)$ are related by complex
conjugation. The longitudinal and transverse currents then acquire no
nontrivial relative phase, strongly suppressing the Stokes parameter
$S_3=2\operatorname{Im}\!\left[J_x^*(\omega)J_y(\omega)\right]$.
For a fixed magnetic domain, TRS is broken, and neither the shift phase
nor $\Delta\mathrm{TDP}$ retains the corresponding antisymmetry between
the paired trajectories. Their contributions are therefore no longer
constrained to be complex conjugates, allowing a finite phase difference
between the longitudinal and transverse harmonic currents. Consequently, this yields a finite $S_3$ and elliptically polarized harmonic emission
(see Supplemental Material, Sec.~7 for details).

In the weak-field regime, the transverse linear current is much smaller
than the dominant longitudinal response. Because both components scale
linearly with the driving-field amplitude, their ratio, and hence the
linear Kerr rotation and ellipticity, is field independent. Nonlinear
effects become important as the field increases. In the
inversion-symmetric system considered here, the leading nonlinear
correction at the fundamental frequency is third order. The complex
ratio $\eta=J_y/J_x$ can therefore be expanded as
$\eta\simeq\eta^{(1)}
+\eta^{(3)}|E_x|^2
+O\!\left(|E_x|^4\right)$ in the perturbation regime,
where $\eta^{(3)}$ contains contributions from both the transverse and
longitudinal third-order conductivities (see Supplemental Material,
Sec.~8). If the third-order term dominates the nonlinear correction
to the transverse current while the longitudinal current remains predominantly linear,
the field-induced change in $\eta$, rather than its total value, scales
quadratically with the field. In the small-angle limit, the variation of the rotation angle \(\Delta\theta\simeq
\operatorname{Re}\!\left[\eta^{(3)}\right]|E_x|^2\), and the variation of the ellipticity \(\Delta\epsilon\simeq
\operatorname{Im}\!\left[\eta^{(3)}\right]|E_x|^2\), respectively.
This scaling accounts for the pronounced nonlinear enhancement of the
SOC-induced H1 ellipticity over the field range considered, while the
rotation exhibits a weaker field dependence. At higher fields,
higher-order terms may produce deviations from the leading quadratic
behavior.

The contrasting field dependences with and without SOC further connect
the nonlinear response to its geometric origin. The black curves in
Figs.~\ref{fig:4}(c) and \ref{fig:4}(d) show that, without SOC, the
rotation and ellipticity change only weakly with field, indicating small
nonlinear magneto-optical corrections. 
Consistently, the shift vectors nearly vanish throughout the Brillouin zone in the absence of SOC, apart from isolated phase singularities, such that the geometric-phase contributions from paired trajectories nearly cancel. SOC instead generates finite, strongly momentum-dependent shift vectors, removes their time-reversal antisymmetry, and produces a continuous texture of \(\Delta\mathrm{TDP}\) (see Supplemental Fig.~S6). As the field-driven trajectories sample this momentum-space texture, they acquire noncanceling, field-dependent shift phases. These phases enter the nonlinear expansion of the interband polarization and contribute to the third-order transverse current underlying the \(\eta^{(3)}|E_x|^2\) response. These results identify geometric-phase accumulation, and in particular its shift-phase contribution, as a key microscopic origin of the field-induced nonlinear magneto-optical enhancement.

In summary, we have established a time-reversal helicity-pairing
constraint on strong-field harmonic emission in the subgap and
weak-injection regime. First-principles semiconductor-Bloch-equation
calculations for bilayer $\mathrm{Cr_2Ge_2Te_6}$ show that SOC imprints
the magnetic order on the orbital wave functions, breaks the
geometric-phase pairing, and produces pronounced ED. Under linearly
polarized driving, the same geometric mechanism generates a nonlinear
transverse current and enhances the harmonic rotation and ellipticity,
thereby separating the magnetic response from the anisotropic
crystallographic background. Helicity-resolved harmonic emission thus
provides a complementary nonlinear probe of SOC-coupled magnetic order,
including in centrosymmetric crystals where bulk electric-dipole SHG is
forbidden.

Antiunitary symmetries such as $\mathcal{PT}$ and $C_2\mathcal{T}$ may
also suppress ED. Phase-resolved harmonic measurements could therefore
help distinguish magnetic point-group symmetries, opening a route toward
strong-field magnetic symmetry spectroscopy.

The authors acknowledge support from the National Key R\&D Program of
China (2021YFA1403800, 2022YFA1604301, 2024YFB3816300), National Natural Science Foundation of China (12425411, 12674425, 12434013, U25D8005),
the Natural Science Foundation of Jiangsu Province
(BK20253027) and Fundamental Research Funds for the
Central Universities (WZJC202602002).

\bibliographystyle{apsrev4-2}
\bibliography{references}

\end{document}


\preprint{}
\title{Supplemental Material: Nonlinear Magneto-Optical Probing of Time-Reversal Symmetry Breaking}

\author{Chen Qian}
\affiliation{Institute of Ultrafast Optical Physics, Department of Applied Physics, Nanjing University of Science and Technology, Nanjing 210094, P. R. China}
\affiliation{Institute of Physics, Chinese Academy of Sciences, Beijing National Laboratory for Condensed Matter Physics, Beijing 100190, P. R. China}
\author{Ziwen Wang}
\affiliation{Institute of Ultrafast Optical Physics, Department of Applied Physics, Nanjing University of Science and Technology, Nanjing 210094, P. R. China}
\author{Shicheng Jiang}
\affiliation{Institute of Ultrafast Optical Physics, Department of Applied Physics, Nanjing University of Science and Technology, Nanjing 210094, P. R. China}
\author{Hui Li}
\affiliation{Institute of Ultrafast Optical Physics, Department of Applied Physics, Nanjing University of Science and Technology, Nanjing 210094, P. R. China}
\author{Kun Zhao}
\affiliation{Institute of Ultrafast Optical Physics, Department of Applied Physics, Nanjing University of Science and Technology, Nanjing 210094, P. R. China}
\author{Chao Yu}
\email{chaoyu@njust.edu.cn}\affiliation{Institute of Ultrafast Optical Physics, Department of Applied Physics, Nanjing University of Science and Technology, Nanjing 210094, P. R. China}
\author{Yuan Wan}
\affiliation{Institute of Physics, Chinese Academy of Sciences, Beijing National Laboratory for Condensed Matter Physics, Beijing 100190, P. R. China}
\author{Hongming Weng}
\email{hmweng@iphy.ac.cn}\affiliation{Institute of Physics, Chinese Academy of Sciences, Beijing National Laboratory for Condensed Matter Physics, Beijing 100190, P. R. China}
\affiliation{School of Physics, University of Chinese Academy of Sciences, Beijing 100049, P. R. China}
\affiliation{Songshan Lake Materials Laboratory, Dongguan, Guangdong 523808, P. R. China}
\author{Ruifeng Lu}
\email{rflu@njust.edu.cn}
\affiliation{Institute of Ultrafast Optical Physics,
Department of Applied Physics,
Nanjing University of Science and Technology,
Nanjing 210094, P. R. China}

\date{\today}
\maketitle

\section*{1. CALCULATION METHODS}\label{sec:supp-1}

Our first-principles calculations were performed within the framework of
density functional theory using the Vienna \emph{ab initio} Simulation
Package (VASP) \cite{Kresse1996CMS}. The bilayer
\(\CGT\) structure
consists of two monolayers stacked in the AB configuration. Structural
relaxations and self-consistent electronic-structure calculations were
carried out using the generalized-gradient approximation with the
Perdew--Burke--Ernzerhof exchange-correlation functional. The on-site
Coulomb interaction of the Cr \emph{3d} electrons was treated using the
GGA+U approach, with an effective Hubbard parameter of
U\textsubscript{eff} value of 1.0 eV on the Cr sites. Long-range van der
Waals (vdW) interactions were included through the DFT-D2 correction
scheme. A vacuum layer of \(15~\text{\AA}\) was introduced along the out-of-plane
direction to avoid spurious interactions between periodic images. The
structural relaxation was performed using a \(12\times12\times1\) Monkhorst-Pack
\(\mathbf{k}\)-point mesh, a plane-wave cutoff energy of 400 eV, and a force
convergence criterion of \(1~\mathrm{meV}/\text{\AA}\). After structural optimization, the
interlayer vdW spacing of the bilayer was found to be approximately
\(3.364~\text{\AA}\), and the optimized in-plane lattice constant was \(6.904~\text{\AA}\).

Maximally localized Wannier functions were constructed using Wannier90
and WannierTools \cite{Mostofi2008CPC,Wu2018CPC}, with the Cr \emph{d}, Ge \emph{p}, and Te
\emph{p} orbitals chosen as the initial projections. The resulting
Wannier-interpolated Hamiltonian provides an efficient and
gauge-consistent framework for evaluating the transition dipole matrix
elements required in simulations \cite{Silva2019PRB}. For the time-dependent
calculations, the Brillouin zone was sampled on a dense \(650\times650\)
\(\mathbf{k}\)-point mesh to accurately resolve the band dispersion and
wave-function variations. The MLWF construction yields an orthonormal
Bloch representation of the target bands in a smooth Wannier gauge,
thereby suppressing numerical discontinuities between neighboring
\(\mathbf{k}\) points and enabling stable evaluation of the dipole matrix
elements and subsequent time propagation.

Numerical calculations were performed by solving the semiconductor Bloch
equations (SBEs) in the instantaneous Bloch representation. The time
evolution of the SBEs was computed using the fourth-order Runge--Kutta
algorithm with a time step of \(0.15~\mathrm{a.u.}\). The resulting
time-dependent density matrix contains both the carrier populations and the
interband coherences. The total current was then decomposed into interband
and intraband contributions as
\begin{equation}
\mathbf{J}_{\mathrm{inter}}(t)
=
-\frac{1}{N_{\mathbf K}}
\sum_{\mathbf K\in\mathrm{BZ}}
\sum_{n\neq m}
\rho_{nm} [\mathbf{k}(t),t]
\mathbf{p}_{nm}\bigl[\mathbf{k}(t)\bigr],
\label{S1}
\end{equation}
\begin{equation}
\mathbf{J}_{\mathrm{intra}}(t)
=
-\frac{1}{N_{\mathbf K}}
\sum_{\mathbf K\in\mathrm{BZ}}
\sum_n
\rho_{nn} [\mathbf{k}(t),t]
\mathbf{p}_{nn}\bigl[\mathbf{k}(t)\bigr].
\label{S2}
\end{equation}
Here, \(N_{\mathbf K}\) is the total number of sampled crystal momenta $\mathbf K$, and the time-dependent crystal momentum is given by \(\mathbf{k}(t)=\mathbf K+\mathbf A(t)\) in atomic units. $\rho_{nm}[\mathbf{k}(t),t]
=\langle a_n^*(\mathbf K,t)a_m(\mathbf K,t)\rangle$
denotes the density matrix in the instantaneous Bloch basis. The velocity
operator is defined as \(\hat{\mathbf p}(\mathbf k)=\nabla_{\mathbf k}\hat{H}(\mathbf k)\),
and its instantaneous Bloch-basis matrix elements are
\(\mathbf{p}_{nm}(\mathbf k)
=
\left\langle
u_{n\mathbf k}
\left|
\hat{\mathbf p}(\mathbf k)
\right|
u_{m\mathbf k}
\right\rangle\).
The off-diagonal and diagonal matrix elements are evaluated as
\begin{equation}
\mathbf{p}_{nm}\bigl[\mathbf{k}(t)\bigr]
=
i\varepsilon_{nm}\bigl[\mathbf{k}(t)\bigr]
\mathbf{d}_{nm}\bigl[\mathbf{k}(t)\bigr],
\qquad n\neq m,
\label{S3}
\end{equation}
\begin{equation}
\mathbf{p}_{nn}\bigl[\mathbf{k}(t)\bigr]
=
\nabla_{\mathbf k}
\varepsilon_n\bigl[\mathbf{k}(t)\bigr].
\label{S4}
\end{equation}
where
\(\varepsilon_{nm}=\varepsilon_n
-\varepsilon_m\), and
\(\mathbf{d}_{nm}(\mathbf k)
=
i\left\langle
u_{n\mathbf k}
\left|
\nabla_{\mathbf k}u_{m\mathbf k}
\right.
\right\rangle\)
is the interband dipole matrix element. Berry-curvature-induced anomalous motion is associated
with field-induced interband mixing and is therefore encoded in the
off-diagonal coherences and velocity matrix elements in the full quantum
description. The intraband current arises from the band group velocity. Under time reversal, the intraband velocities change
sign accordingly. In the weak-injection regime, where the populations remain approximately paired, the intraband current obeys the same symmetry relation as the interband current, $\mathbf{J}_{n,n}[\epsilon,\mathbf{M};\mathbf{k}(t),t]=-\mathbf{J}_{\tilde{n}\,\tilde{n}}[-\epsilon,-\mathbf{M};-\mathbf{k}(-t),-t]$.

The harmonic spectra were calculated
as the squared modulus of the Fourier-transformed total current. In the absence of SOC, the total harmonic response is obtained by
superposing the contributions from the spin-up and spin-down band
manifolds. Numerical convergence is confirmed using four valence bands
and four conduction bands.

\section*{2. WEAK-INJECTION AND MULTIPHOTON-EXCITATION REGIMES}
\label{sec:supp-2}

\begin{figure}[t]
\centering
\smartinclude[width=0.78\linewidth]{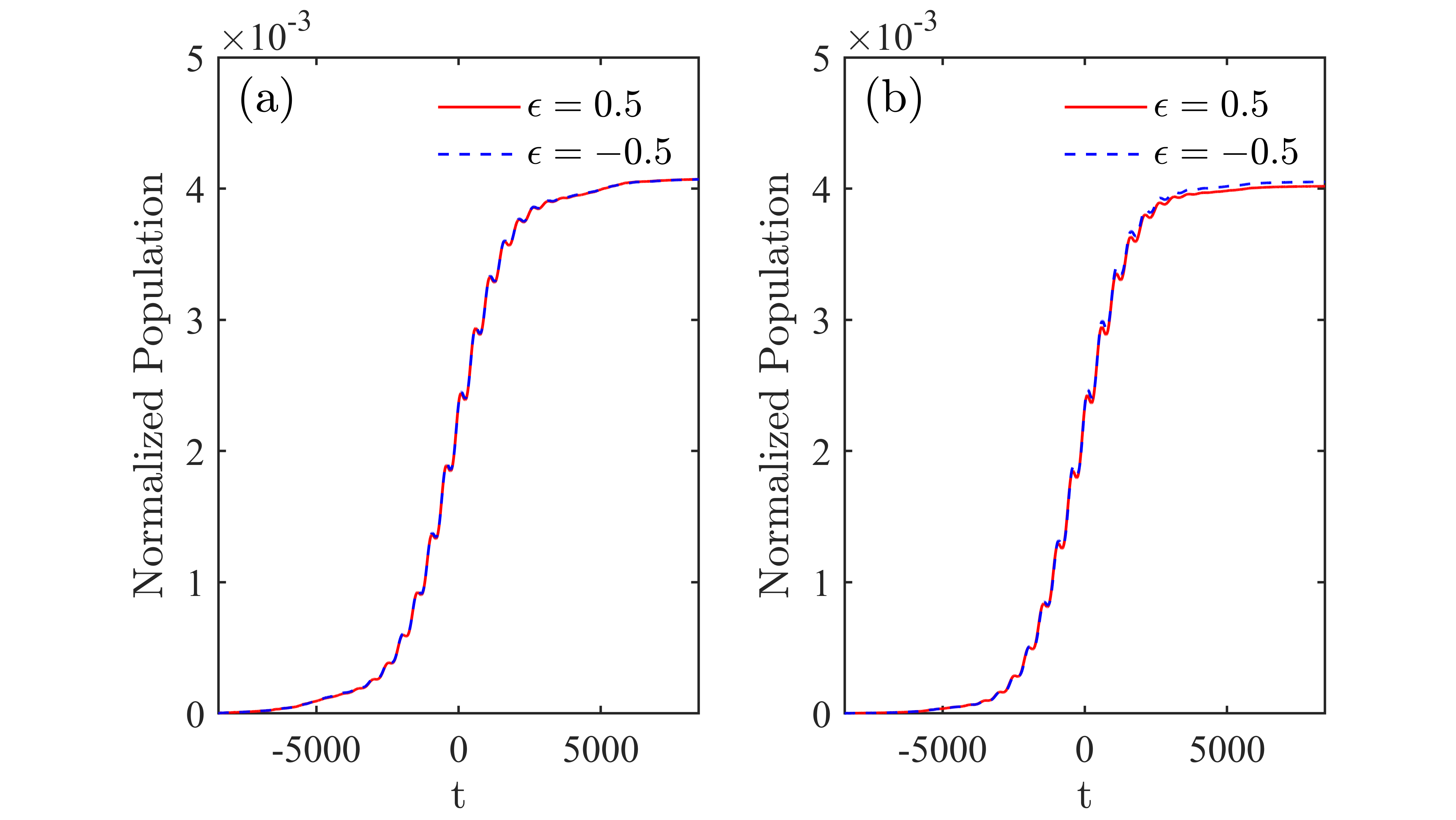}
\caption{Time-dependent electron population in all conduction bands from $\CGT$ (a) without SOC and (b) with SOC. The electric-field amplitude is the same as that in Fig. 3 of the main text. The total population is normalized.}
\label{fig:S1}
\end{figure}

To verify that the system remains in the weak-injection regime, we examine
the time-dependent conduction-band electron population under the driving
field, as shown in Fig.~\ref{fig:S1}. In the absence of spin--orbit coupling
(SOC), the carrier populations generated by fields with opposite
ellipticities remain nearly identical throughout the laser pulse, indicating
the absence of intrinsic elliptical dichroism (ED). When SOC is included, a
small difference gradually emerges between the populations excited by the
two opposite ellipticities. This population difference remains weak because
inversion symmetry constrains the band dispersions and the magnitudes of the
transition-dipole matrix elements at opposite crystal momenta. Therefore, the
substantially stronger ED observed in the emitted high-order harmonics
suggests that the magnetization-dependent asymmetry in the phase of the
induced current plays an essential role. Moreover, the accumulated
conduction-band population remains below $0.5\%$, confirming that the field
strength of $1.54$~MV/cm used in the main calculations lies within the
weak-injection regime.

The excitation mechanism can be further characterized using the Keldysh
parameter,
    \(\gamma =
    \frac{\omega_0 \sqrt{m^{*}E_{\mathrm{g}}}}{eE_0}\).
In the calculations, we use laser frequency $\omega_0 = 7500$ nm ($6.067 \times 10^{-3}$~a.u.), 
effective mass $m^{*} = 0.35m_0$ and
band gap $E_{\mathrm{g}} = 0.636$~eV
($2.337 \times 10^{-2}$~a.u.) for the SOC case. Over the electric-field range considered in
this work, $E_0 = 0.5 \times 10^{-4}$\textasciitilde$4 \times 10^{-4}$~a.u., the Keldysh
parameter decreases from approximately $11.0$ to $1.4$. Thus, $\gamma$
remains larger than unity throughout the entire field range, indicating that
carrier excitation is dominated by multiphoton processes rather than
tunneling ionization. Together with the low conduction-band population, this
analysis confirms that the calculations are performed in the
weak-injection, multiphoton-excitation regime.

\section*{3. MAGNETIZATION-REVERSAL DEPENDENCE OF THE HARMONIC RESPONSE}\label{sec:supp-3}

\begin{figure}[t]
\centering
\smartinclude[width=0.68\linewidth]{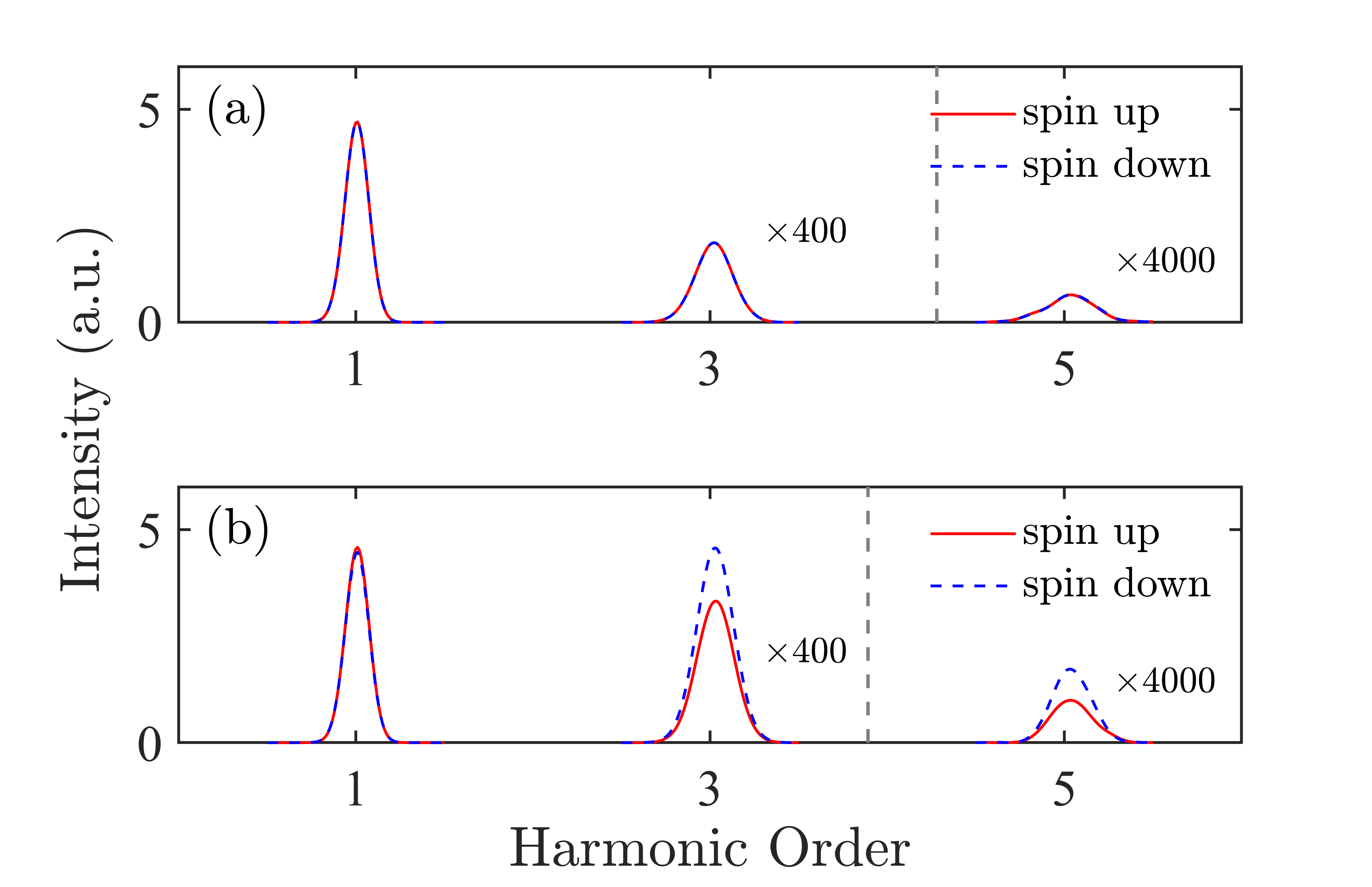}
\caption{Magnetization-reversal dependence of the harmonic response from $\CGT$ (a) without SOC and (b) with SOC. The laser ellipticity $\epsilon=0.5$, other laser parameters are the same as those in Fig. 3 of the main text.}
\label{fig:S2}
\end{figure}

As shown in Fig.~\ref{fig:S2}, when SOC is not included, there is no magnetization-reversal harmonic difference between the spin-up and spin-down polarization cases, indicating that the orbital wave functions still preserve spinless time-reversal symmetry (TRS). Upon the inclusion of SOC, however, the harmonics exhibit a clear magnetization-dependent difference, demonstrating that the magnetic order breaks the TRS of the orbital wave functions.

\section*{4. ELLIPTICAL DICHROISM OF TIME-DOMAIN PHOTOCURRENT}\label{sec:supp-4}

\begin{figure}[t]
\centering
\smartinclude[width=0.78\linewidth]{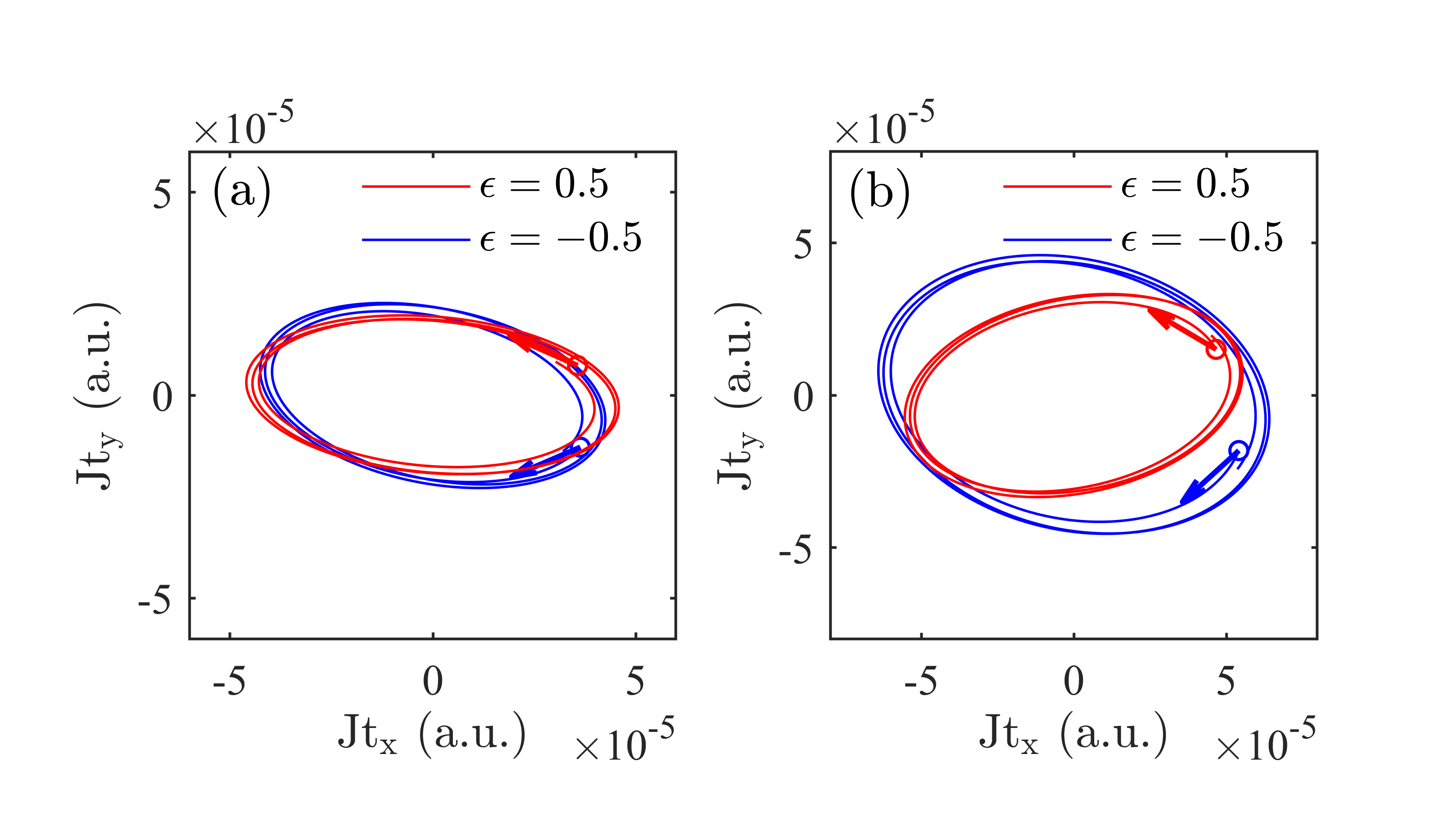}
\caption{In-plane third-order photocurrent of $\CGT$ without SOC (a) and with SOC (b) obtained through bandpass filtering. The directions of current flow over time are denoted by arrows. The laser parameters used are the same as those in Fig. 3 of the main text.}
\label{fig:S3}
\end{figure}

Figure S3 shows the time-domain photocurrent of third-order harmonic (H3) driven by
elliptically polarized light. In the absence of SOC, the third-harmonic
currents driven by fields of opposite ellipticities are nearly identical
after the appropriate time-reversal transformation [Fig.~\ref{fig:S3}(a)] and
approximately
satisfy~\(\mathbf{J}(t,\epsilon) = - \mathbf{J}(-t,-\epsilon)\).
The time interval plotted here is
\(\left\lbrack - \frac{T_{0}}{2},\frac{T_{0}}{2} \right]\),
\(T_{0}\) is a single cycle. When the time is reversed, the photocurrent
will flow along the opposite direction. Once SOC is included, the
currents generated by opposite ellipticities no longer obey this pairing
relation exactly, as shown in Fig.~\ref{fig:S3}(b). The resulting mismatch in the
time-domain trajectories gives rise to the harmonic ED.

\section*{5. DEPENDENCE ON LASER ELECTRIC-FIELD STRENGTH OF ED}\label{sec:supp-5}

\begin{figure}[t]
\centering
\smartinclude[width=0.8\linewidth]{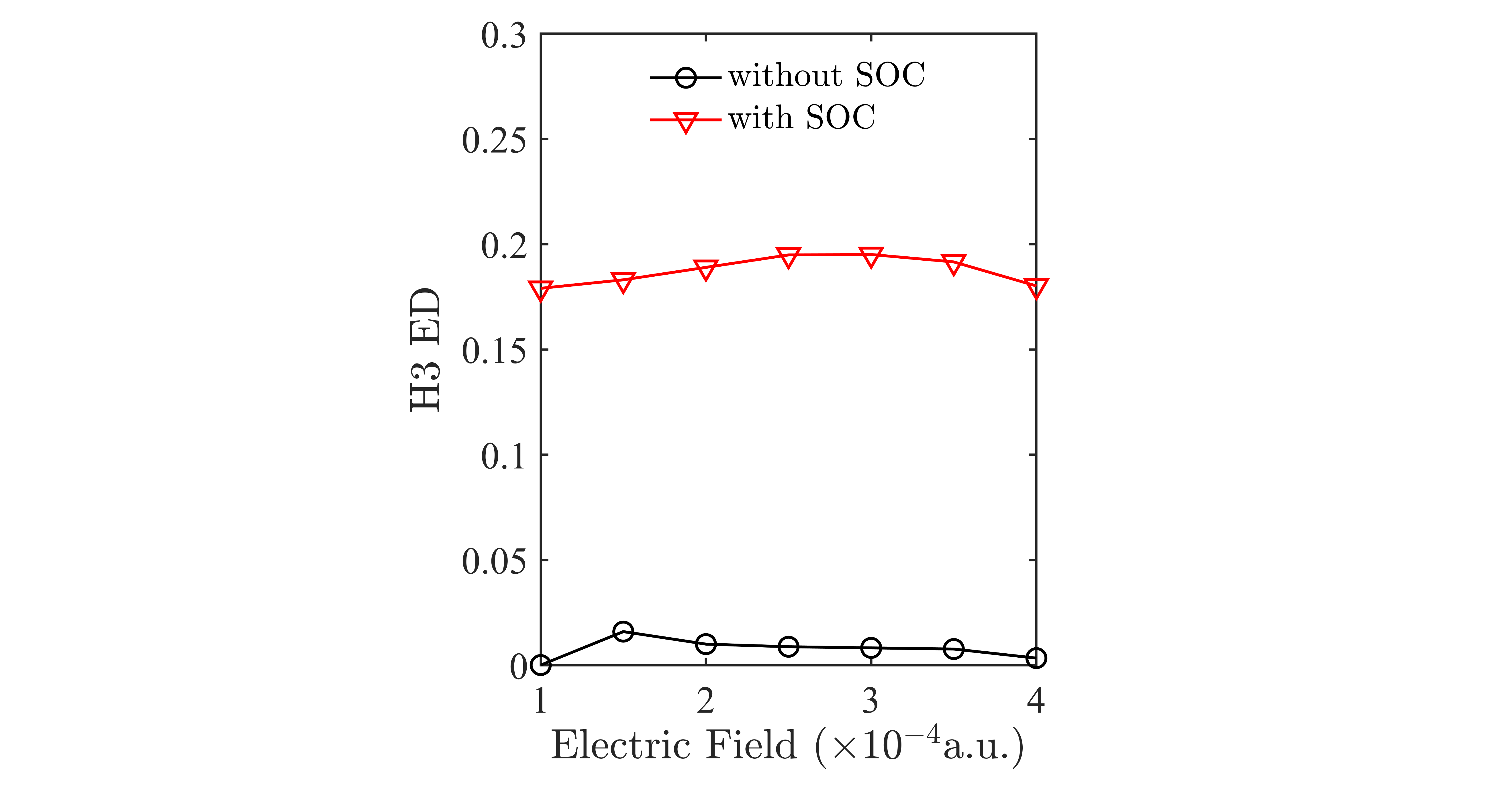}
\caption{ED as a function of electric-field strength for H3 under elliptically polarized driving with ellipticity $\epsilon=\pm0.5$.}
\label{fig:S4}
\end{figure}

In the subgap, weak-injection regime, low-order harmonics are governed
predominantly by reversible polarization currents, whereas carrier
injection and interband charge transfer remain negligible. Under these
conditions, time-reversal-related channels are approximately paired,
leading to a strong suppression of the ED. As shown in Fig.~\ref{fig:S4}, the ED
is nearly zero in the absence of SOC. Upon inclusion of SOC, however,
the ED of H3 becomes pronounced and remains relatively large with
increasing field strength. This result indicates that nonlinear ED is
highly sensitive to time-reversal-symmetry breaking in the orbital wave
functions, thereby revealing the nonlinear character of the strong-field
magneto-optical response.

\section*{6. GLOBAL ELLIPTICITY DEPENDENCE}\label{sec:supp-6}

\begin{figure}[t]
\centering
\smartinclude[width=0.82\linewidth]{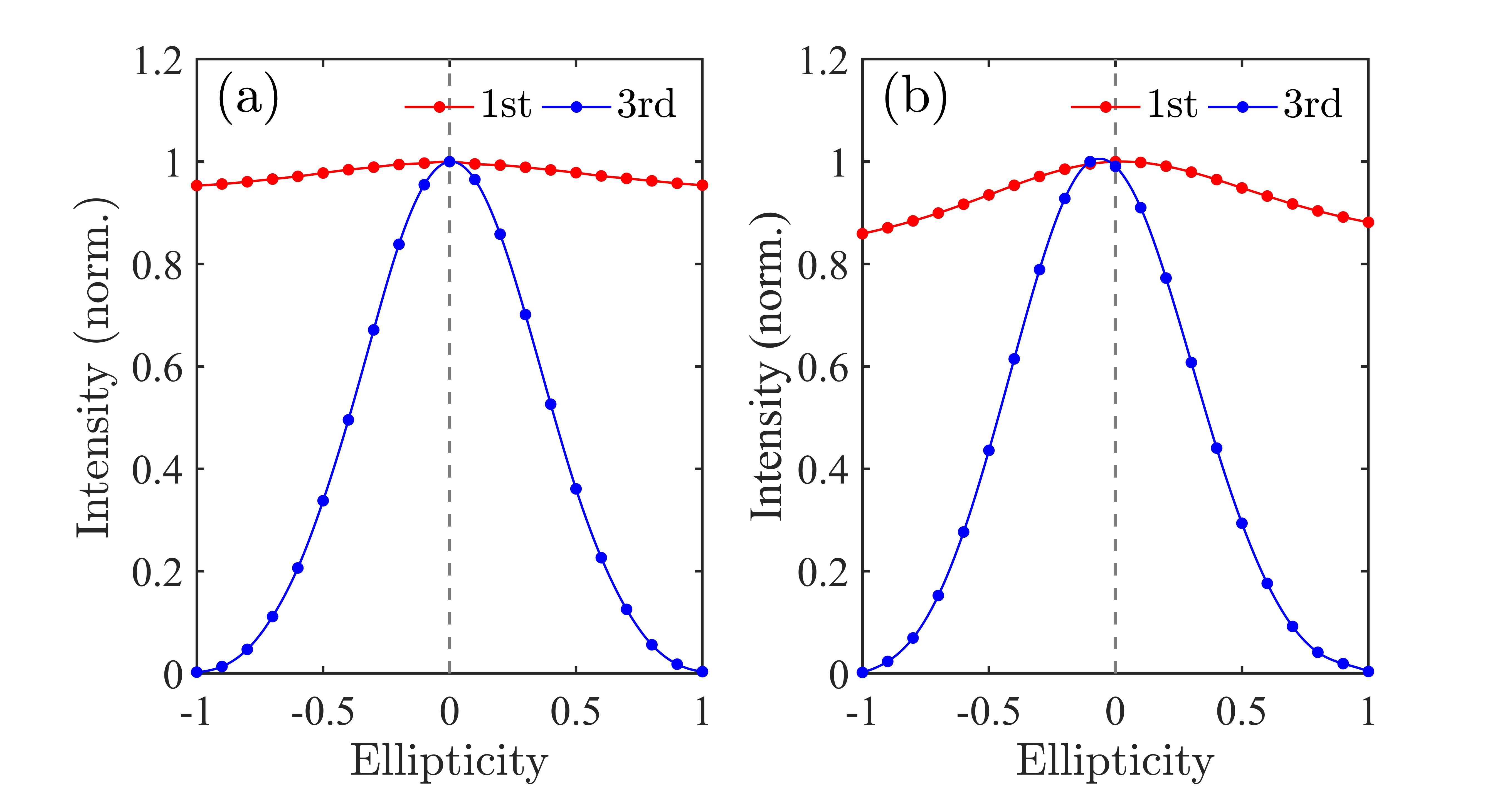}
\caption{Ellipticity dependence of harmonic intensity from $\CGT$ (a) without SOC and (b) with SOC. The other parameters used are the same as those in Fig. 3 of the main text.}
\label{fig:S5}
\end{figure}

In Fig.~\ref{fig:S5}, we show the global dependence of harmonic intensity on
the ellipticity of the driving laser. For the H1 and H3, their intensities
are symmetric with respect to the laser ellipticity \(\epsilon = 0\),
when SOC is not considered. After taking into account the SOC, the
broken TRS in the spin space is coupled to the orbital wave functions.
Through the electric dipole effect, the symmetry breaking can be
reflected in the elliptical dichroism of the harmonics. Our discussion
is not limited to H3, but applies to the harmonics below the
bandgap energy. The 5th harmonic is slightly above the bandgap, so it
will also deviate from the symmetric behavior without SOC. If symmetry
allows, the 2nd harmonic is also within the scope of our theory.

\section*{7. ELLIPTICITY UNDER LINEARLY POLARIZED DRIVING}\label{sec:supp-7}

In this paper, the rotation angle and ellipticity of harmonic radiation
are defined using the Stokes parameters. The polarization state of the
harmonic radiation can be directly defined using the frequency-domain
current, which can be written as
\begin{equation}\mathbf{J}(\omega)=J_x(\omega)\hat{\mathbf{x}}+J_y(\omega)\hat{\mathbf{y}},\label{eq:S3}\end{equation}
where \(J_{x}(\omega)\) and \(J_{y}(\omega)\) are complex
frequency-domain current amplitudes.

The Stokes parameters are defined as
\begin{equation}\begin{aligned}S_0&=|J_x|^2+|J_y|^2,\\S_1&=|J_x|^2-|J_y|^2,\\S_2&=2\operatorname{Re}[J_x^*J_y],\\S_3&=2\operatorname{Im}[J_x^*J_y].\end{aligned}\label{eq:S4}\end{equation}
The corresponding rotation angle, namely the angle by which the
principal polarization axis is rotated with respect to the
\(x\)-direction, is
\begin{equation}\theta=\frac{1}{2}\operatorname{atan2}(S_2,S_1).\label{eq:S5}\end{equation}
The ellipticity is defined as
\begin{equation}\epsilon=\tan\!\left[\frac{1}{2}\arcsin\!\left(\frac{S_3}{S_0}\right)\right].\label{eq:S6}\end{equation}
Therefore, whether the harmonic radiation is elliptically polarized is
determined by \(S_{3}\), namely by
\(\operatorname{Im}\left[ J_{x}^{*}(\omega)J_{y}(\omega) \right]\). If
\(\operatorname{Im}\left[ J_{x}^{*}(\omega)J_{y}(\omega) \right] = 0\),
then \(J_{x}(\omega)\) and \(J_{y}(\omega)\) are in phase or out of
phase, and the radiation is linearly polarized. If it is nonzero, there
exists a nontrivial phase difference between the two orthogonal
components, and the radiation has finite ellipticity.

\subsection*{7.1 Time-reversal symmetric case}\label{sec:supp-7-1}

As stated in the main text, the time-reversal operation acting on the
elliptically polarized current component satisfies
\begin{equation}\hat{\mathcal T}:\mathbf{J}(\epsilon,\mathbf{M};t)\rightarrow -\mathbf{J}(-\epsilon,-\mathbf{M};-t) = \mathbf{J}(\epsilon,\mathbf{M};t).\label{eq:S7}\end{equation}
This relation shows that, for linearly polarized driving, the current is
odd in a time-reversal-symmetric crystal. In other words, for the Cartesian components,
\begin{equation}J_a(-t)=-J_a(t),\qquad a=x,y.\label{eq:current relation}\end{equation}

Now consider the Fourier component at the fundamental harmonic. Since
\(J_{a}(t)\) is a real odd function, it can be expanded using sine
Fourier components:
\begin{equation}J_a(\omega)=-\frac{i}{T}\int_{-T/2}^{T/2}J_a(t)\sin(\omega t)\,dt,\label{eq:S9}\end{equation}
where \(T\) is an integer number of optical periods or an effective time
window of the pulse. Thus,
\(J_{x}^{*}(\omega)J_{y}(\omega) \in \mathbb{R}\). According to the
definition of the Stokes parameters,
\begin{equation}S_3=2\operatorname{Im}[J_x^*(\omega)J_y(\omega)]=0.\label{eq:S10}\end{equation}
This shows that, in a time-reversal symmetric system, linearly polarized
driving does not generate a net elliptically polarized fundamental
harmonic. Equivalently, the ellipticity should be approximately zero.

\subsection*{7.2 Time-reversal broken case}\label{sec:supp-7-2}

\begin{figure*}[t]
    \centering
    \includegraphics[width=\textwidth]{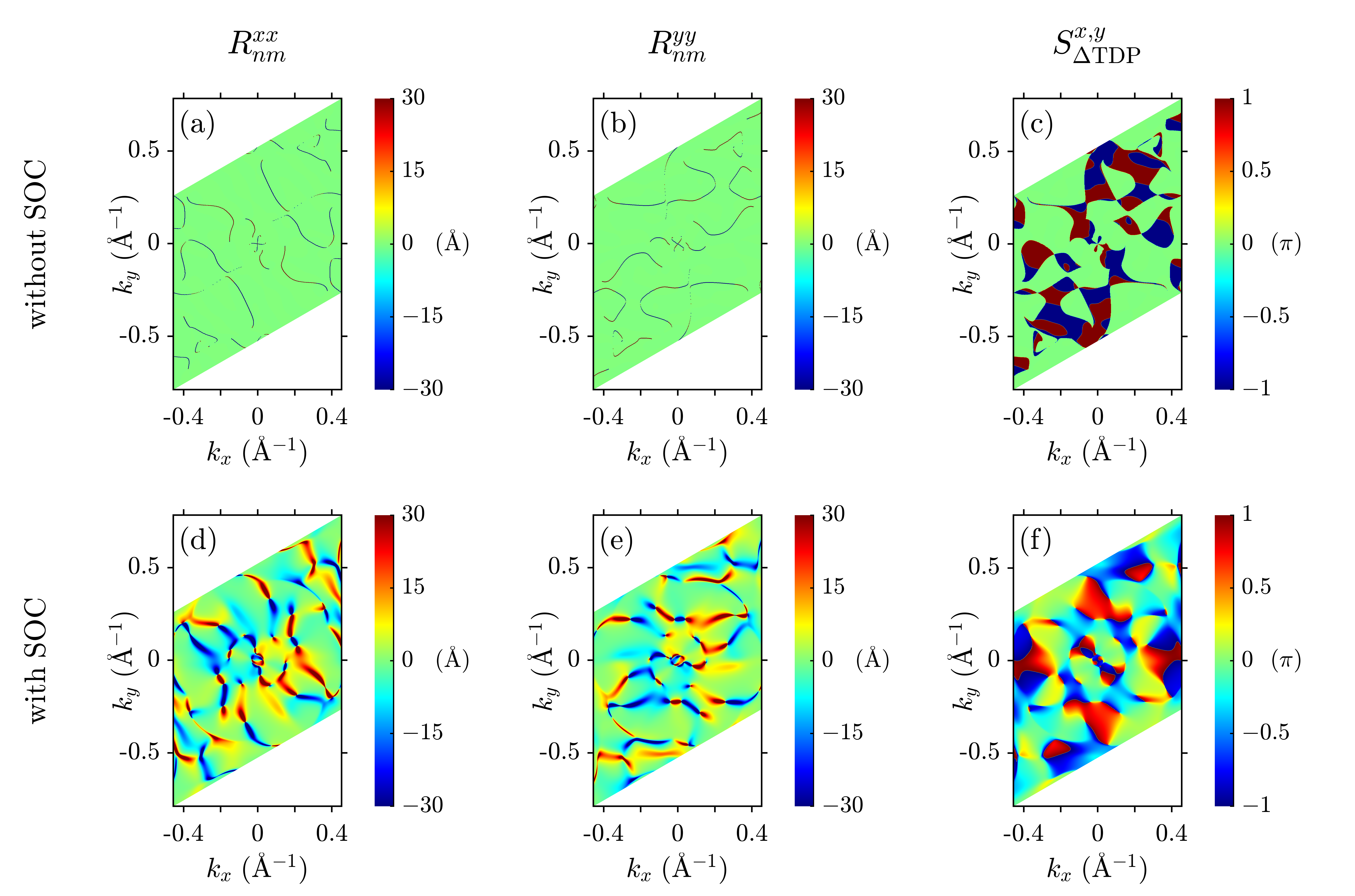}
    \caption{
    Momentum-space distributions of the geometric quantities entering the
    interband polarization current in bilayer
    $\mathrm{Cr_2Ge_2Te_6}$.
    (a),(b) Shift-vector components $R_{nm}^{xx}$ and $R_{nm}^{yy}$ between the highest valence band and the lowest conduction band,
    respectively, and (c) transition-dipole phase difference
    $S_{\Delta\mathrm{TDP}}^{x,y}$, calculated without SOC.
    The shift vectors vanish almost everywhere, except at isolated phase
    singularities, while $\Delta{\mathrm{TDP}}$ is restricted to
    $0$ and $\pm\pi$.
    (d)--(f) Corresponding quantities calculated with SOC for a fixed
    magnetization. SOC produces finite, strongly momentum-dependent shift
    vectors and a continuous texture of $\Delta{\mathrm{TDP}}$, thereby
    removing the time-reversal antisymmetry of the geometric phase.
    The shift vectors are given in units of \AA, and
    $\Delta{\mathrm{TDP}}$ is given in units of $\pi$.
    }
    \label{fig:S6}
\end{figure*}

In a magnetic system, time reversal reverses the magnetic order:
\begin{equation}\hat{\mathcal T}:\ \mathbf{M}\rightarrow-\mathbf{M}.\label{eq:S11}\end{equation}
Therefore, for a fixed magnetic domain~\(\mathbf{M}\), the system itself
no longer satisfies TRS. In this case, the phase
relation between
the~\((\mathbf{k},t)\) and~\(( -\mathbf{k},-t)\) trajectories
is no longer strictly complex conjugate.

Microscopically, the phase of the strong-field current mainly includes
the dynamical phase, the shift phase, and the transition-dipole-moment phase difference ($\Delta\mathrm{TDP}$). Schematically, it can be written as
\begin{equation}
S=S_{\mathrm{dyn}}+S_{\mathrm{shift}}+S_{\Delta\mathrm{TDP}}.
\label{eq:S12}
\end{equation}
For an electron--hole pair created at time $t'$ and subsequently evolved 
to time $t$, and 
$\mathbf{k}(\tau)$ denotes its instantaneous crystal momentum. The shift 
phase accumulated along this trajectory is
\begin{equation}
S_{\mathrm{shift}}^{b,b}(\mathbf{k},t,t')
=\int_{t'}^{t}
E^{b}(\tau)\cdot
R_{nm}^{b,b}[\mathbf{k}(\tau)]\,\mathrm{d}\tau,
\label{eq:Sshift}
\end{equation}
where $E^{b}(\tau)$ is the driving field along the $b$ direction 
and
\begin{equation}
R_{nm}^{b,b}(\mathbf{k})
=d_{nn}^{b}(\mathbf{k})
-d_{mm}^{b}(\mathbf{k})
-\partial_{k_b}\phi_{nm}^{b}(\mathbf{k})
\label{eq:Rshift}
\end{equation}
is the corresponding shift vector. Here, $d_{nn}^{b}$ and 
$d_{mm}^{b}$ are the intraband Berry connections of bands $n$ 
and $m$, respectively, while $\phi_{nm}^{b}$ is the phase of the 
transition dipole moment defined through
$d_{nm}^{b}(\mathbf{k})
=|d_{nm}^{b}(\mathbf{k})|
e^{i\phi_{nm}^{b}(\mathbf{k})}$.
The shift vector characterizes the relative displacement 
of the charge centers of the two bands. 
Finally, the $\Delta\mathrm{TDP}$ between the $a$ and $b$ directions is defined as
\begin{equation}
S_{\Delta\mathrm{TDP}}^{a,b}[\mathbf{k}(t)]
=\phi_{nm}^{a}[\mathbf{k}(t)]-\phi_{nm}^{b}[\mathbf{k}(t)],
\label{eq:DeltaTDP}
\end{equation}
where $\phi_{nm}^{a}$ and $\phi_{nm}^{b}$ denote the phases of the transition dipole moments along the photocurrent direction $a$ and the laser-field direction $b$, respectively. The $\Delta\mathrm{TDP}$ therefore provides an additional phase difference between the polarization-resolved current components, thereby contributing to the polarization properties of the harmonic emission.

In the absence of SOC, the spin and orbital sectors are decoupled, and each
spin-resolved orbital Hamiltonian obeys the effective complex-conjugation
relation
\begin{equation}
    H_s(\mathbf{k})=H_s^{*}(-\mathbf{k}),
\end{equation}
even though the spin-polarized system as a whole breaks time-reversal
symmetry. Together with inversion symmetry, this orbital
time-reversal-like relation strongly constrains the geometric quantities
entering the interband polarization current. As shown in
Figs.~\ref{fig:S6}(a)-\ref{fig:S6}(c), the shift-vector components vanish throughout the
Brillouin zone, except at isolated phase singularities where they are not
well defined, while the transition-dipole phase difference takes only the
quantized values
$\Delta{\mathrm{TDP}}=0,\pm\pi$. Because $\pi$ and $-\pi$ are equivalent
modulo $2\pi$, the geometric-phase distribution is compatible with both
the inversion-even and time-reversal-odd constraints. Consequently, the
geometric phases $S_{\mathrm{geo}}=S_{\mathrm{shift}}+S_{\Delta\mathrm{TDP}}$ accumulated along the two time-reversal-paired
trajectories are opposite modulo $2\pi$:
\begin{equation}
    S_{\mathrm{geo}}(\mathbf{k},t,t')
    =
    -S_{\mathrm{geo}}(-\mathbf{k},-t,-t')
    \quad (\mathrm{mod}\ 2\pi)
    \label{eq:geometric symmetry}
\end{equation}
This relation ensures that the current relation in Eq.~(\ref{eq:current relation}) is recovered
after summation over the Brillouin zone.

When SOC is included, however, the orbital wave functions no longer obey
the above complex-conjugation relation within a fixed magnetic domain.
Inversion symmetry continues to constrain the band energies and hence
protects the corresponding symmetry of the dynamical phase, but it does
not restore the time-reversal antisymmetry of the geometric phases. This
distinction is directly visible in Figs.~\ref{fig:S6}(d)-\ref{fig:S6}(f): the shift vectors
become finite and strongly momentum dependent, and
$\Delta{\mathrm{TDP}}$ develops a continuous momentum-space texture
instead of being restricted to $0$ and $\pm\pi$. In particular, the
resulting geometric-phase distribution no longer satisfies Eq.~(\ref{eq:geometric symmetry}) for a fixed magnetization.

Under inversion, the shift vector is odd,
\begin{equation}
    \mathbf{R}_{nm}(\mathbf{k},\mathbf{M})
    =
    -\mathbf{R}_{nm}(-\mathbf{k},\mathbf{M}).
\end{equation}
Because inversion also reverses the direction of the field-driven
momentum-space trajectory, the shift phase accumulated along a pair of
inversion-related trajectories is inversion-even rather than
time-reversal-odd. Likewise, the transition-dipole phase contribution
$\Delta{\mathrm{TDP}}$ is inversion-even. These even-parity geometric
contributions are therefore not constrained to satisfy the
antisymmetry required by time reversal. Namely, for a fixed magnetic
domain, one generally obtains
\begin{equation}S_{\mathrm{shift}}(\mathbf{k},t,t',\mathbf{M})\neq-S_{\mathrm{shift}}(-\mathbf{k},-t,-t',\mathbf{M}).\label{eq:S13}\end{equation}
 and
\begin{equation}S_{\Delta\mathrm{TDP}}(\mathbf{k},t,t',\mathbf{M})\neq-S_{\Delta\mathrm{TDP}}(-\mathbf{k},-t,-t',\mathbf{M}).\label{eq:S14}\end{equation}
Therefore, the
contributions
from~\((\mathbf{k},t)\) and~\(( -\mathbf{k},-t)\) are
no longer strictly complex conjugates. As a result, a nontrivial phase
difference can arise between the transverse and longitudinal currents.
Thus,
\begin{equation}S_3=2\operatorname{Im}[J_x^*(\omega)J_y(\omega)]\neq0.\label{eq:S16}\end{equation}
According to the Stokes-parameter definition, a
finite~\(S_{3}\)~directly corresponds to finite ellipticity.
Therefore, the generation of elliptically polarized fundamental harmonic
radiation in a magnetic system essentially originates from the breaking
of the symmetry of the shift phase and \(\Delta\mathrm{TDP}\) after TRS is broken.

\section*{8. LINEAR KERR EFFECT AND NONLINEAR STRONG-FIELD ENHANCEMENT}\label{sec:supp-8}

\subsection*{8.1 Linear Hall current derived from \MainEqOne}\label{sec:supp-8-1}

In the weak-field limit, the field-driven trajectory can be approximated
as \(\mathbf{k}(t) \approx \mathbf{K}\), and the population difference
is replaced by its equilibrium
value \(f_{nm}\left( \mathbf{K},t \right) = \rho_{nn}\left( \mathbf{K},0 \right) - \rho_{mm}\left( \mathbf{K},0 \right)\). Since \MainEqOne\ already contains a term linear in the electric
field,~\(\mathbf{E}\left( t^{'} \right)\mathbf{\cdot}\left| \mathbf{d}_{mn}\left( \mathbf{K} \right) \right|\),
the shift
phase~\(\int_{t^{'}}^{t}{\mathbf{E}(\tau)\mathbf{\cdot}\mathbf{R}_{nm}\left\lbrack \mathbf{k}(\tau) \right]}d\tau\)~contributes
only to second and higher orders in the field and is therefore neglected
in the linear response regime. Fourier transformation of \MainEqOne\ then
yields
\begin{equation}J_a^{(1)}(\omega)=\sigma_{ab}^{\mathrm{inter}}(\omega)E_b(\omega).\label{eq:S17}\end{equation}
where
\begin{equation}\sigma_{ab}^{\mathrm{inter}}(\omega)=-\frac{i}{N_{\mathbf K}}\sum_{\mathbf{K}\in\mathrm{BZ}}\sum_{n\neq m}\frac{\varepsilon_{nm}(\mathbf{K})d_{nm}^{a}(\mathbf{K})d_{mn}^{b}(\mathbf{K})f_{nm}(\mathbf{K})}{\omega-\varepsilon_{nm}(\mathbf{K})+i\gamma}, \qquad \gamma \rightarrow 0^{+}.\label{eq:S18}\end{equation} 
This expression corresponds to the interband
part of the standard Kubo formula for the optical conductivity. In the low-frequency or dc limit, the antisymmetric
off-diagonal component of the optical conductivity reduces to the Hall
conductivity. Specifically,
\begin{equation}J_a^{\mathrm{H}}=\sigma_{ab}^{\mathrm{H}}E_b.\label{eq:S19}\end{equation}
the transverse Hall conductivity
\(\sigma_{ab}^{\mathrm{H}} = \frac{1}{2}\left\lbrack \sigma_{ab}^{\text{inter}}(\omega) - \sigma_{ba}^{\text{inter}}(\omega) \right]\).
For an intrinsic anomalous Hall response, this low-frequency or dc Hall
conductivity can be expressed in terms of the Berry curvature as
\begin{equation}\sigma_{ab}^{\mathrm{H}}=-\frac{1}{N_{\mathbf K}}\sum_{\mathbf{K}\in\mathrm{BZ}}\sum_n f_n(\mathbf{K})\Omega_n^{ab}(\mathbf{K}).\label{eq:S20}\end{equation}
with
\begin{equation}\Omega_n^{ab}(\mathbf{K})=-2\operatorname{Im}\sum_{m\neq n}\frac{p_{nm}^{a}(\mathbf{K})p_{mn}^{b}(\mathbf{K})}{[\varepsilon_m(\mathbf{K})-\varepsilon_n(\mathbf{K})]^2}.\label{eq:S21}\end{equation}
Thus, the transverse first-order optical current obtained from \MainEqOne\ continuously connects to the conventional linear Hall current in the
low-frequency limit.

For a two-dimensional insulating state with a vanishing total Chern number of the occupied bands, the intrinsic dc Hall conductivity vanishes. In the weak-field linear-response regime, when the band gap is much larger than the driving photon energy, the low-frequency antisymmetric off-diagonal optical response of a topologically trivial magnetic insulator is generally strongly suppressed and is typically much weaker than the dominant longitudinal polarization response. Therefore, under strong-field driving, nonlinear response mechanisms can become the dominant contribution to the transverse current response.

Within the interband-polarization-current framework employed here, such
nonlinear contributions mainly originate from field-induced changes in
the carrier population, the dynamical evolution of carriers in momentum
space, and geometric phase effects, including the shift phase. Since the
longitudinal polarization current is generally dominated by the linear
response, polarimetry of strong-field harmonic emission, such as
harmonic elliptical dichroism, polarization rotation, and ellipticity,
provides a sensitive probe of the nonlinear growth of the transverse
polarization current.

\subsection*{8.2 Independence of the linear Kerr effect on electric-field strength}\label{sec:supp-8-2}

In the limit of \(\left| J_{x} \right| \gg \left| J_{y} \right|\),
corresponding to both the rotation angle and ellipticity being small
quantities here, the polarization state can be characterized by the
complex ratio
\(\eta(\omega) \equiv \frac{J_{y}(\omega)}{J_{x}(\omega)}\). The Kerr
rotation angle and ellipticity are approximately
\begin{equation}\theta\simeq\operatorname{Re}(\eta),\qquad\epsilon\simeq\operatorname{Im}(\eta).\label{eq:S22}\end{equation}

The ratio between the transverse and longitudinal linear response
currents is
\begin{equation}\eta^{(1)}=\frac{J_y^{(1)}}{J_x^{(1)}}=\frac{\sigma_{yx}^{(1)}}{\sigma_{xx}^{(1)}},\label{eq:S23}\end{equation}
where \(\sigma_{xx}^{(1)}\) is the longitudinal linear optical
conductivity, and \(\sigma_{yx}^{(1)}\) is the transverse linear optical
conductivity allowed in a magnetic system. Since both \(J_{x}^{(1)}\)
and \(J_{y}^{(1)}\) are proportional to the driving field amplitude
\(E_{0}\), their ratio is independent of \(E_{0}\).

The linear magneto-optical rotation and ellipticity may therefore be
finite, but they do not systematically increase with the driving-field
strength. In other words, if the calculated or experimentally measured
harmonic polarization rotation or ellipticity increases as the electric-field
strength increases, then the enhanced part cannot be explained by the
purely linear Kerr effect. Instead, it should originate from the
nonlinear strong-field response.

\subsection*{8.3 Nonlinear fundamental harmonic response under inversion symmetry}\label{sec:supp-8-3}

If the system has spatial inversion symmetry, even-order electric-dipole
responses are forbidden, and the fundamental harmonic current can
contain only odd-order responses. For the fundamental harmonic current
at frequency \(\omega_{0}\), one can write
\begin{equation}\begin{aligned}J_a(\omega_0;E_b)&=\sigma_{ab}^{(1)}E_b(\omega_0)+\sigma_{abbb}^{(3),\mathrm{eff}}E_b(\omega_0)E_b(\omega_0)E_b^*(\omega_0)+\cdots,\\[-2pt]&\qquad a,b=x,y,\end{aligned}\label{eq:S24}\end{equation}
where \(\sigma_{ab}^{(1)}\) is the linear conductivity, while
\(\sigma_{abbb}^{(3),\mathrm{eff}}\) is effective third-order nonlinear
conductivity. Therefore, the third-order response can contribute to the
fundamental harmonic.
Thus, the ratio of longitudinal and transverse fundamental harmonic
currents is
\begin{equation}\eta(\omega_0;E_x)=\frac{\sigma_{yx}^{(1)}+\sigma_{yxxx}^{(3),\mathrm{eff}}E_x(\omega_0)E_x^*(\omega_0)+\cdots}{\sigma_{xx}^{(1)}+\sigma_{xxxx}^{(3),\mathrm{eff}}E_x(\omega_0)E_x^*(\omega_0)+\cdots}.\label{eq:S25}\end{equation}

When the third-order term provides the dominant nonlinear current and the longitudinal current remains dominated by the linear response, the ratio can be expanded as
\begin{equation}\eta(\omega_0;E_x)\simeq\eta^{(1)}+\eta^{(3)}|E_x(\omega_0)|^2+O(|E_x|^4).\label{eq:S26}\end{equation}
where
\begin{equation}\eta^{(3)}=\frac{\sigma_{yxxx}^{(3),\mathrm{eff}}-\eta^{(1)}\sigma_{xxxx}^{(3),\mathrm{eff}}}{\sigma_{xx}^{(1)}}.\label{eq:S27}\end{equation}
Here, \(\eta^{(1)} = \frac{\sigma_{yx}^{(1)}}{\sigma_{xx}^{(1)}}\)
corresponds to the linear Kerr background, while \(\eta^{(3)}\)
corresponds to the leading nonlinear term.

Therefore, for a magnetic system that already exhibits a finite linear Kerr effect, strong-field driving can produce a substantial quadratic nonlinear enhancement of the Kerr signal on top of the finite linear-response background.

In the main text, we show that \(\theta(E_0)\) or \(\epsilon(E_0)\) increases significantly as \(E_0\) increases; this indicates that both the real and imaginary combinations
of the frequency-domain current have been modified by nonlinear
strong-field effects.

\bibliographystyle{apsrev4-2}
\bibliography{references}